\documentclass[aps,pra,twocolumn,showpacs,floatfix]{revtex4-2}
\usepackage{graphicx}
\usepackage{amsmath}
\usepackage{amssymb}
\usepackage{color}
\usepackage[separate-uncertainty = true,multi-part-units=single]{siunitx}
\newcommand{\Hefour}[0]{$^4\mathrm{He}$}
\newcommand{\HeII}{He\,II}

\begin{document}

\title{Controlled excitation of rotons in superfluid helium with an optical centrifuge}

\author{Alexander~A.~Milner and Valery~Milner}

\affiliation{Department of  Physics \& Astronomy, The University of British Columbia, Vancouver, Canada}

\date{\today}

\begin{abstract}
We experimentally demonstrate a controlled transfer of angular momentum to roton pairs in superfluid helium. The control is executed with an optical centrifuge and detected with coherent time- and frequency-resolved Raman scattering. We show that the sign of the Raman shift, and hence the orientation of the angular momentum transferred from the laser field to the rotons, is dictated by the centrifuge. The magnitude of the shift reflects the two-roton energy and indicates that the centrifuge-induced hot roton pairs are far from the equilibrium with the colder quantum bath. The observed decay of the coherent Raman signal suggests that the decoherence is governed by the scattering on thermal rotons and phonons. The demonstrated method offers ways of examining microscopic origins of superfluidity by controlling collective excitations in superfluids.
\end{abstract}
\maketitle

Superfluid helium is one of the paradigm systems for studying collective excitations in strongly interacting many-body quantum systems. As first proposed by Landau \cite{Landau1941a,Landau1941b}, elementary excitations in superfluid \Hefour{} (\HeII{}) consist of phonons, maxons and rotons. Many aspects of superfluidity can be described and explained by the dispersion properties of these quasi-particles and their interactions. Until recently, they have been studied predominantly with experimental tools sensitive to the equilibrium steady-state dynamics of the superfluid -- neutron scattering \cite{Palevsky1958,Yarnell1959} and spontaneous Raman scattering \cite{Greytak1969,Greytak1970} (for a recent review, see Ref.~\citenum{Glyde2017}). The latter approach followed a theoretical prediction that rotons can be optically observed only in pairs, owing to the big momentum mismatch with a much smaller momentum of a photon \cite{Halley1969}.

Spectral broadening of the single-roton peak in experiments with neutrons \cite{Andersen1994a,Andersen1994b,Gibbs1999}, and the two-roton peak in Raman scattering \cite{Greytak1970,Ohbayashi1998} has been associated with roton-roton and roton-phonon collisions in the gas of quasi-particles in thermal equilibrium \cite{Iwamoto1970,Ruvalds1970,Zawadowski1972,Bedell1984}. Recently implemented time-resolved optical study of roton pairs revealed rich non-equilibrium dynamics with both the central frequency and spectral width of the two-roton peak changing on a picosecond time scale during the thermalization of rotons with the superfluid bath \cite{Milner2023a}.

In addition to the energy shift and linewidth of scattered light, its polarization provides further insight into the nature of collective excitations in superfluid helium \cite{Stephen1969}. Since rotons are bosons, the angular momentum $\ell$ of a roton pair can attain only even values \cite{Bedell1984}. Polarization analysis of the two-roton Raman spectra of \HeII{} showed that the scattering is predominantly of $d$-wave character, which stems from the ``dipole--induced-dipole'' interaction mechanism \cite{Udagawa1986}: The incident light polarizes helium atoms, whose fluctuating density (governed by phonons and rotons) results in an electric field at the neighboring volume of other He atoms; the induced oscillating polarization of the latter is responsible for the emission of the scattered light. Experimentally observed $d$-wave scattering of light by a roton pair implied an exchange of two units of angular momentum \cite{Udagawa1986,Shay2007}.

Whereas \textit{spontaneous} Raman scattering samples thermal roton pairs with isotropically distributed vectors of angular momentum $\boldsymbol\ell$, we utilize \textit{stimulated} Raman scattering to excite two-roton states in a controlled way with a well-defined orientation of $\boldsymbol\ell$. To that end, we employ an optical centrifuge -- a unique laser tool previously used for the controlled rotational excitations of molecules \cite{Karczmarek1999,Villeneuve2000,Yuan2011,Korobenko2014a}, to control the transfer of angular momentum from the laser field to the collective excitations in a superfluid.

We detect the rotational state of a roton pair by means of coherent Raman scattering (CRS) of weak probe laser pulses. Owing to its polarization sensitivity \cite{Korobenko2014a, Milner2014a}, CRS probing not only indicates that the centrifuge-induced two-roton excitation carries two units of angular momentum (as inferred from earlier experiments), but also shows that its orientation is dictated by the direction of rotation of the optical centrifuge. Femtosecond time resolution of the coherent Raman technique enables us to study the decay of the laser-induced two-roton angular momentum and compare it to the lifetime of the roton pair.

\begin{figure*}[t]
  \includegraphics[width=.9\textwidth]{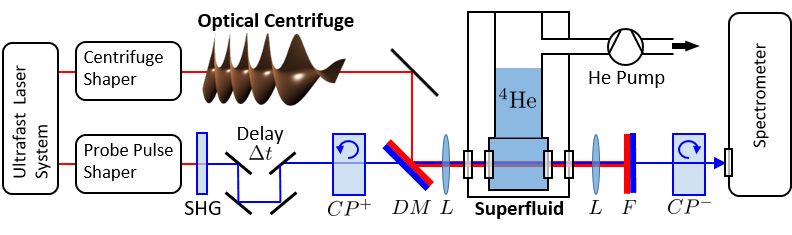}\\
  \caption{Diagram of the experimental setup. Femtosecond pulses from the ultrafast laser system are used for creating the field of an optical centrifuge, schematically illustrated by the corkscrew shape (top), and for generating time-delayed frequency-doubled circularly polarized probe pulses (bottom). The centrifuge and the probe beams are combined in a collinear geometry and focused into the bulk liquid helium, kept at the variable temperature and saturated vapor pressure. The spectrum of the probe pulses is recorded with a spectrometer. $SHG$: second harmonic generation crystal for frequency doubling, $CP^{\pm}$: circular polarizers of opposite handedness, $DM$: dichroic mirror, $L$: lens, $F$: spectral filter.}
  \label{fig-Setup}
\end{figure*}
The experimental setup is illustrated in Fig.\ref{fig-Setup}. Femtosecond pulses from a Ti:Sapphire laser system ($\SI{35}{fs}$ pulse length, $\SI{1}{KHz}$ repetition rate, $\SI{793}{nm}$ central wavelength) are shaped into the field of an optical centrifuge -- a laser pulse whose linear polarization rotates with accelerating rate \cite{Karczmarek1999}. Our setup for producing the centrifuge has been described in a recent review \cite{MacPhail2020}. Briefly, we split the spectrum of each pulse in two equal parts using the original recipe of Villeneuve \textit{et al.} \cite{Villeneuve2000}. The two spectral parts are then frequency chirped in opposite directions and circularly polarized with an opposite handedness. When combined together, interference of such laser fields results in the rotation of the polarization vector with an angular frequency growing linearly in time at the constant rate of \SI{\approx 100}{GHz/ps}. As will be shown below, a \SI{\approx 4}{ps}-long centrifuge pulse, used in this work, sweeps through the energy of a two-roton state (\SI{\approx 360}{GHz}) and leads to its efficient excitation.

The centrifuge pulses ($\lesssim40$~$\mu$J pulse energy, $\lesssim1.5\times10^{12}$~W/cm$^2$ peak intensity) are focused in the bulk liquid helium, condensed in a custom-built optical cryostat. By pumping the helium gas from the cryostat, the temperature of the liquid can be varied between $\SI{1.4}{K}$ and $\SI{4.2}{K}$ at the respective saturated vapor pressure (SVP) above the liquid's surface. Each centrifuge pulse is followed by a weak probe pulse, derived from the same laser system and spectrally narrowed down to the bandwidth of \SI{\approx 100}{GHz} (pulse length of \SI{4.4}{ps}) in a standard $4f$ Fourier pulse shaper \cite{Weiner2011}. Frequency doubling of probe pulses shifts their central wavelength to \SI{398}{nm}, which allows an easy separation from the centrifuge beam with a spectral filter.

Probe pulses are circularly polarized at the entrance to the cryostat by a circular polarizer (a combination of a linear polarizer and a quarter-wave plate, $CP^{+}$ in Fig.~\ref{fig-Setup}), and passed through a circular analyzer of an opposite handedness ($CP^{-}$) after exiting the sample. This choice of an input/output probe polarization stems from the conservation of angular momentum in a parametric four-wave mixing process of Raman scattering, in which one circularly polarized photon from the centrifuge field is absorbed, while another photon of opposite circular polarization is returned back into the centrifuge field via stimulated emission. As a result, two units of angular momentum are exchanged between the laser field and the medium, while coherence is induced between the participating angular momentum states. Owing to the latter, the polarization handedness of the scattered light is opposite to the input probe polarization \cite{Korobenko2014a}. The probe spectrum is recorded with a \SI{100}{GHz}(\SI{4.8}{K})-resolution spectrometer as a function of the centrifuge-probe delay time, the temperature of the liquid and the direction of the centrifuge rotation.

\begin{figure}[b]
  \includegraphics[width=1\columnwidth]{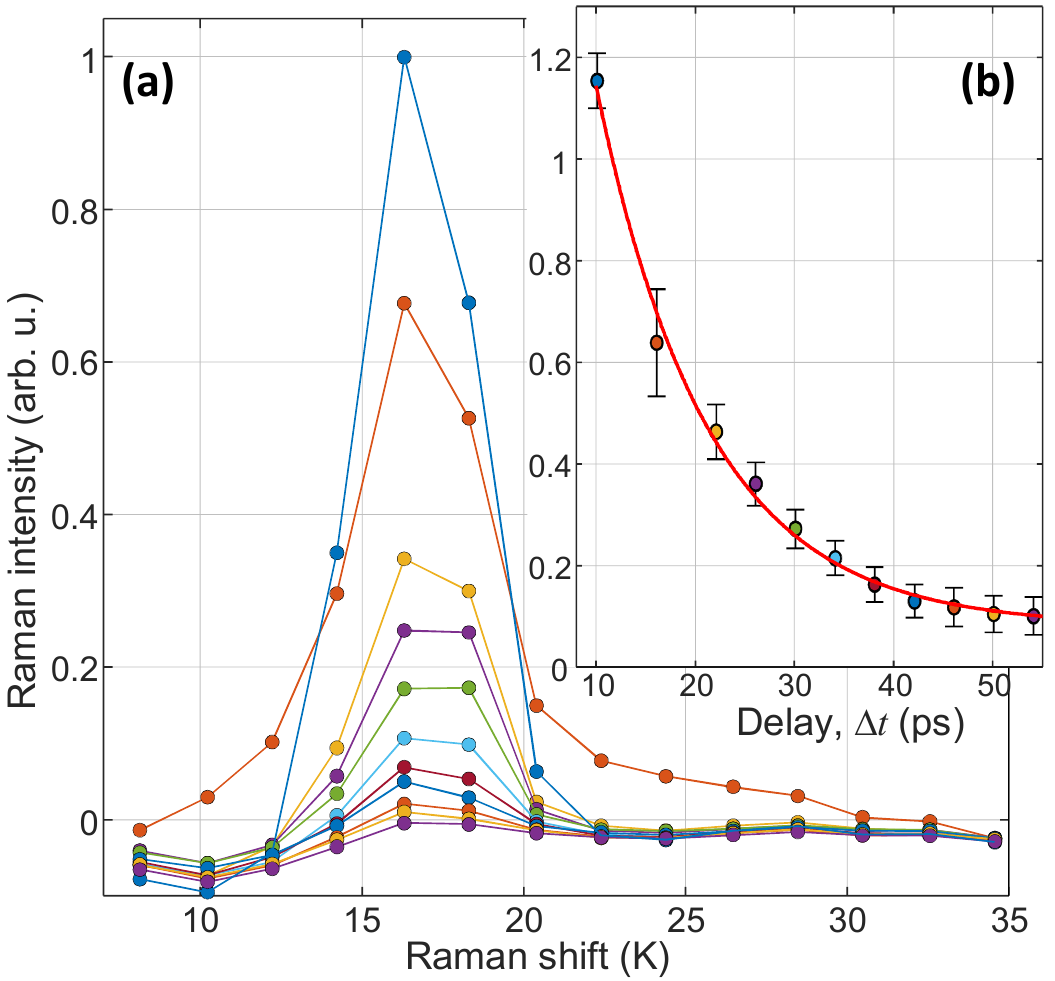}\\
  \caption{(\textbf{a}) Coherent Raman spectra of the superfluid helium at $T=\SI{1.37(2)}{K}$ and respective SVP under the excitation by the optical centrifuge. From the top down, the spectra were recorded at different delay times between the centrifuge and probe pulses, from $\Delta t=\SI{10}{ps}$ to \SI{54}{ps}. (\textbf{b}) Dependence of the Raman peak amplitude on $\Delta t$. The fit to an exponential decay is depicted with the solid red line.}
  \label{fig-RamanTime}
\end{figure}
\begin{figure*}[t]
  \includegraphics[width=.75\textwidth]{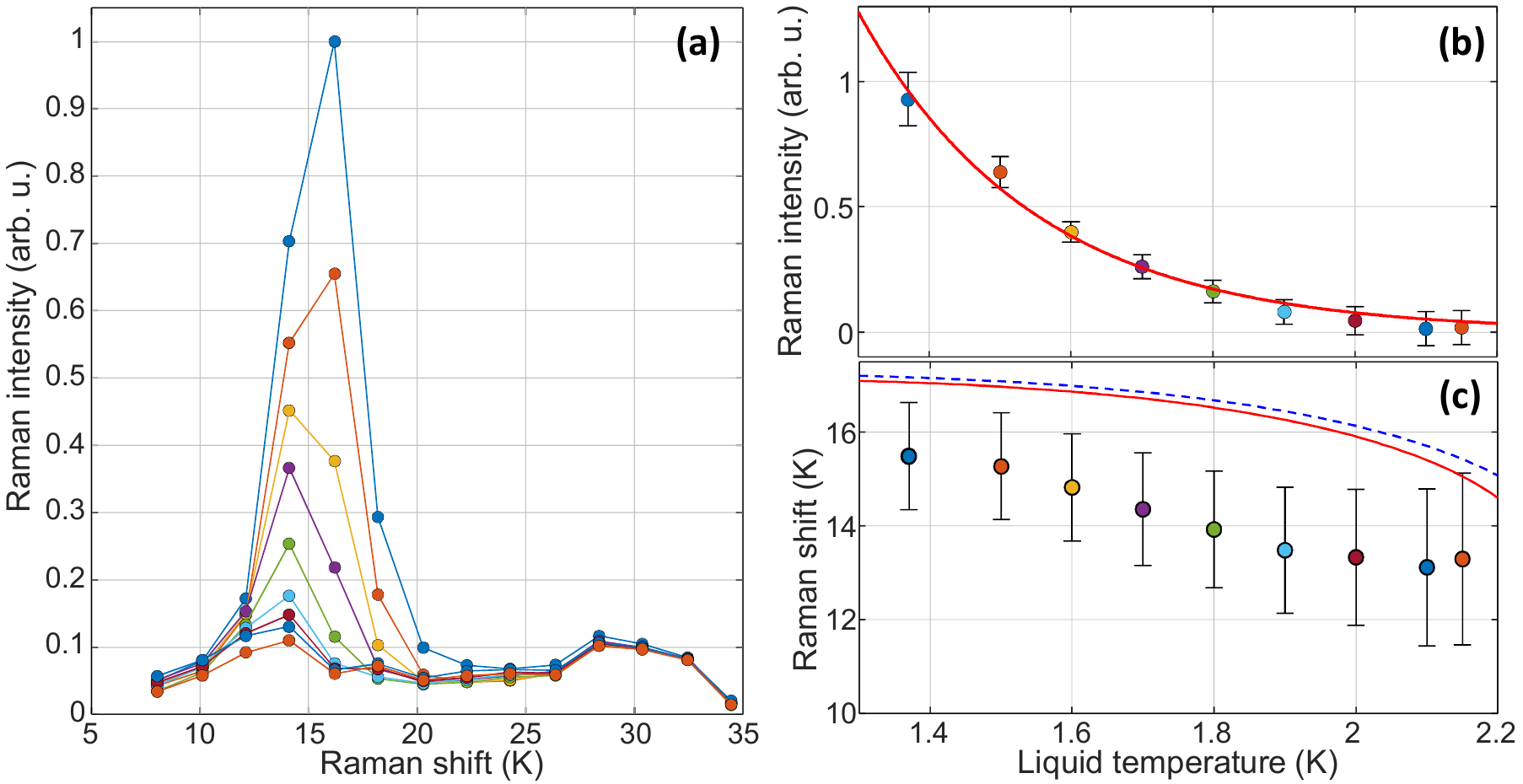}\\
  \caption{(\textbf{a}) Coherent Raman spectra of superfluid helium under the centrifuge excitation at the centrifuge-probe delay time $\Delta t=\SI{6}{ps}$. From the top down, the helium temperature was increased from $T=\SI{1.37(2)}{K}$ to \SI{2.15(2)}{K}. (\textbf{b}) Dependence of the peak amplitude on $T$. The fit to an exponential dependence is plotted with the solid red line. (\textbf{c}) Raman shift as a function of temperature. Solid red (dashed blue) lines depict the $T$-dependence of the two-roton energy found in the neutron scattering experiments \cite{Andersen1996} (\cite{Pearce2001}).}
  \label{fig-RamanTemperature}
\end{figure*}
Figure~\ref{fig-RamanTime}(\textbf{a}) shows a set of probe spectra taken at the constant liquid helium temperature of \SI{1.37(2)}{K} and different delays between the centrifuge and probe pulses. The two-roton Raman peak appears when the probe pulse is scanned in time across the \SI{4}{ps}-long centrifuge, and decays thereafter. We define zero delay ($\Delta t=0$) as the time at which the Raman signal at the two-roton frequency reaches its maximum. Strong non-resonant Raman background, owing to the instantaneous electronic response of the liquid to the excitation fields, dictates the minimum delay of $\gtrsim\SI{4}{ps}$. Here, we started the scan at $\Delta t=\SI{10}{ps}$ and normalized the recorded spectra to the highest peak at the shortest delay. All spectra show a clear Raman line shifted from the central probe wavelength by $\SI{17(1)}{K}$, which agrees well with the energy of a two-roton state at this temperature \cite{Halley1969, Greytak1969}. The accuracy of the energy shift and the measured line width are both limited by the resolution of our spectrometer.

Increasing the delay from \SI{10}{ps} to \SI{54}{ps} results in the exponential drop in the peak amplitude. This is presented in panel (\textbf{b}) of Fig.~\ref{fig-RamanTime}. The amplitude of each Raman line was calculated by fitting it to a Gaussian profile. The decay of the Raman signal is well described by an exponent with a time constant of \SI{11.2\pm1.7}{ps} [solid red line in Fig.~\ref{fig-RamanTime}(\textbf{b})]. This time is noticeably shorter than a two-roton \textit{equilibrium} lifetime of $\approx\SI{36}{ps}$ at $T=\SI{1.37}{K}$, obtained from both the neutron scattering \cite{Pearce2001} and the spontaneous Raman data \cite{Ohbayashi1998}. It indicates the underlying \textit{non-equilibrium} dynamics, similar to that found in our recent work on roton pairs excited with a linearly polarized femtosecond pulse \cite{Milner2023a}.

There, the two-roton energy increased from \SI{16.90}{K} to \SI{17.05}{K} in the same time window between 10 and \SI{50}{ps} (not visible in the Raman shifts in Fig.~\ref{fig-RamanTime} due to the limited spectral resolution). Since roton energy increases with decreasing temperature of the superfluid \cite{Bedell1982,Bedell1984}, it was attributed to the ultrafast cooling of initially ``hot'' roton pairs by the colder bath. Noteworthy is the very similar time constant of \SI{11.6\pm0.5}{ps} for the energy relaxation found in that study and the CRS amplitude decay observed here. It suggests that the loss of angular momentum by a roton pair is governed by the same mechanism as its equilibration with the quantum environment, i.e. by scattering on thermal rotons and phonons.

From the fit to the exponential decay in Fig.~\ref{fig-RamanTime}(\textbf{b}), we find that the amplitude of the Raman signal approaches a non-zero asymptotic value of $(3.4\pm 2.1)\%$ relative to its extrapolated strength at $\Delta t=0$. We attribute this result to the temperature dependent decay constant, which becomes longer as the centrifuge-excited roton pairs cool down. The effect appears similar to an almost two-fold decrease of the roton linewidth observed on the same time scale in our previous work \cite{Milner2023a}.

The dependence of the Raman spectra on the temperature of helium at a fixed delay of \SI{6}{ps} is shown in Fig.~\ref{fig-RamanTemperature}(\textbf{a}). The spectra were normalized to the highest peak, corresponding to the lowest temperature of \SI{1.37(2)}{K}. As we increased the temperature of the liquid to \SI{2.15(2)}{K}, the two-roton Raman line weakened and shifted to lower energies. The decrease of the amplitude of the Raman peaks is plotted in panel (\textbf{b}) of Fig.~\ref{fig-RamanTemperature}, where it is fitted to an exponential dependence $\propto\exp\left[-T/T_0 \right]$ with $T_0=\SI{0.25\pm0.04}{K}$ (solid red line). To within the accuracy of our method, the coherent Raman signal reaches zero at the lambda point. This result is consistent with the two-fluid model, according to which the roton excitations disappear at $T_{\lambda }$ together with the superfluid component \cite{Woods1978, Andersen1994b, Glyde2017}. Unfortunately, taking data at $T>T_{\lambda}$ proved impossible due to the laser-induced thermal instabilities in the liquid above the phase transition.

The influence of the \HeII{} temperature on the Raman shift, and hence on the energy of the roton pairs excited by the optical centrifuge, is shown in Fig.~\ref{fig-RamanTemperature}(\textbf{c}). Comparing our data with the results of the neutron scattering experiments \cite{Andersen1996,Pearce2001} (solid red and dashed blue lines, respectively), we again notice the signs of non-equilibrium dynamics. Lower than expected (for a given temperature of \HeII{}) energy of a roton pair indicates that it is substantially hotter than the surrounding medium, suggesting that the rotons did not have enough time to equilibrate with the bath. The observed two-roton energy is lower than in Fig.~\ref{fig-RamanTime}, owing to the shorter centrifuge-probe delay and correspondingly farther-from-equilibrium state of roton pairs. A shorter delay also explains the residual non-resonant (and hence, temperature independent) Raman peak around \SI{30}{K} in Fig.\ref{fig-RamanTemperature}(\textbf{a}).

\begin{figure}[t]
  \includegraphics[width=0.8\columnwidth]{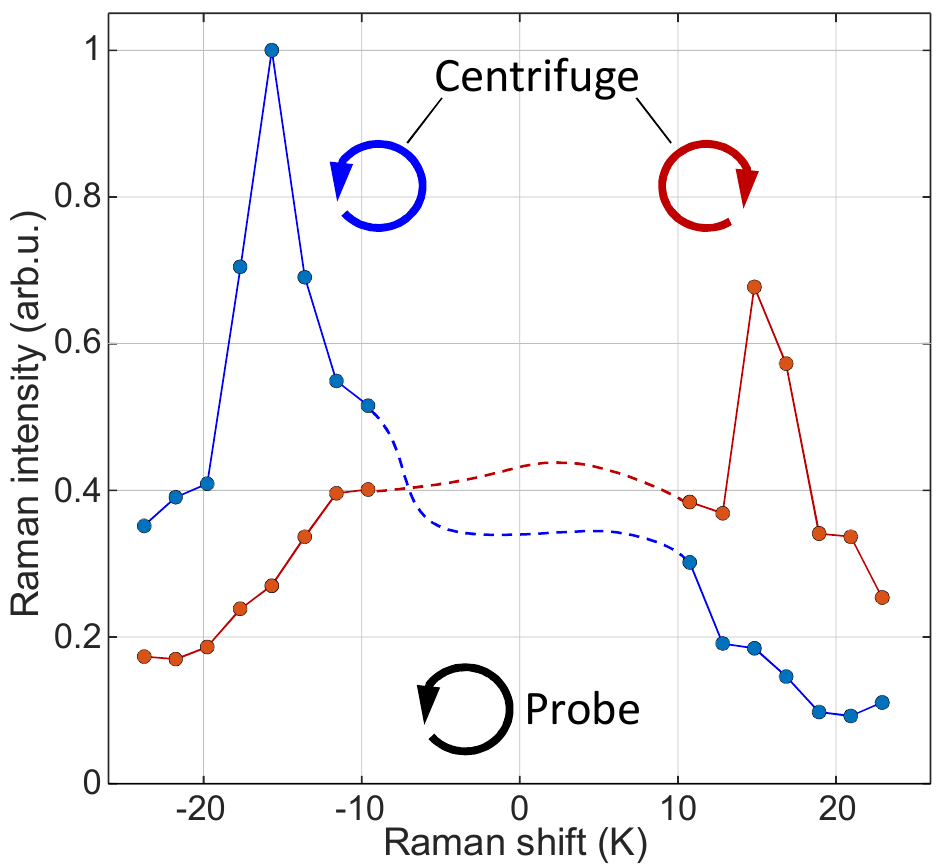}\\
  \caption{Coherent Raman spectra recorded with two opposite directions of rotation of the centrifuge field, indicated at the top. The data was taken at $T=\SI{1.36(2)}{K}$ and $\Delta t=\SI{4}{ps}$. The polarization rotation of the circularly polarized probe field is in the counter-clockwise direction. A much stronger Rayleigh peak at \SI{0}{K} has been removed for clarity, and the two dashed lines were added to connect the opposite sides of the same spectrum.}
  \label{fig-Directionality}
\end{figure}
Our ability to control the two-roton excitations with the optical centrifuge is illustrated in Fig.~\ref{fig-Directionality}. Here, the two coherent Raman spectra were recorded under identical conditions, except from the direction of rotation of the centrifuge field, indicated near the respective Raman peak. The Stokes peak, down-shifted from the central frequency of the probe pulse by the two-roton frequency (blue line), is observed when the polarization vector of the centrifuge is rotating in the same direction as the circular polarization of the probe. Changing the direction of the centrifuge rotation to the opposite results in the appearance of the anti-Stokes peak (red line).

The observed effect of rotational directionality on coherent Raman scattering is very similar to what has been seen in numerous experiments on centrifuged molecules (for a recent review, see Ref.~\citenum{MacPhail2020}). At any given time, the centrifuge pulse supplies two photons of opposite circular polarization, one shifted above and one below its central frequency. An absorption of the higher-energy photon, together with a stimulated emission of the lower-energy one, results in the transfer of two units of angular momentum to the centrifuged system. In the experiments with centrifuged molecules, repeating this process $N$ times makes the molecule climb the rotational ladder of states higher and higher in multiple steps of $2\hbar$. At the end, the molecular angular momentum increases by $\Delta J=2\hbar N$, while its direction is determined by the handedness of the centrifuge field.

In the case studied here, the predominant $d$-wave scattering of light from a two-roton state found in the earlier experiments on spontaneous Raman, suggests that the interaction between the optical centrifuge and the superfluid helium should terminate at the very first Raman step with $\Delta l=2\hbar$. This would result in a single Raman line, in agreement with what we observe and in contrast to the comb of multiple rotational Raman lines typical for a free-rotating molecular systems \cite{Korobenko2014a}. Yet similarly to the molecular case, the handedness of the centrifuge field is transferred to the two-roton state and is carried by that collective excitation for more than \SI{50}{ps}, long after the centrifuge pulse is gone.

In summary, we experimentally demonstrate a controlled transfer of angular momentum form the laser field to two-roton excitation in superfluid helium. The transfer is executed with an optical centrifuge and monitored with coherent Raman scattering. We show that roton pairs excited by the centrifuge retain information about the direction of the centrifuge rotation for $\gtrsim\SI{50}{ps}$, at least an order of magnitude longer than the centrifuge pulse. The decay rate of the induced angular momentum, determined through the decay of coherent Raman signal, is very similar to the recently found rate at which laser-excited hot rotons equilibrate with the colder environment \cite{Milner2023a}. Thus, both processes can be associated with the collisions of the centrifuged roton pairs with thermal rotons and phonons.

The combination of time and frequency resolution offered by the CRS technique allowed us to measure the two-roton energy at short delays after the excitation pulse, down to \SI{4}{ps}, much shorter than in our previous time-resolved study, where a scan of at least \SI{20}{ps} had to be carried out for determining an average roton energy in that broad time window. Our results suggest that shortly after the laser pulse, the excited roton pairs may be much further from the equilibrium than what can be inferred from a single exponent fitted at later times. The observed shifts of roton energy at short delays indicate their effective temperature comparable with (and possibly even higher than) $T_{\lambda }$, almost a full degree above the temperature of the superfluid bath.

Applying an optical centrifuge to inject angular momentum in \HeII{}, and using coherent optical methods to track the centrifuge-induced collective excitations, offers a way of studying them in strongly interacting quantum systems far from equilibrium. Improving the spectral resolution of our detection scheme, which is currently underway, will enable us to reveal further details of these non-equilibrium coherent dynamics of rotons in superfluid helium.

\section*{Acknowledgments}
This research was supported by the Natural Sciences and Engineering Research Council of Canada (NSERC).

%

\end{document}